\documentclass[apj]{emulateapj}
\usepackage{apjfonts}

\usepackage{amsmath,natbib}

\makeatletter

\newcommand{\Rmnum}[1]{\expandafter\@slowromancap\romannumeral #1@}
\makeatother

\newcommand{\gps}{\ensuremath{g_{\rm P1}}}
\newcommand{\rps}{\ensuremath{r_{\rm P1}}}
\newcommand{\ips}{\ensuremath{i_{\rm P1}}}
\newcommand{\zps}{\ensuremath{z_{\rm P1}}}
\newcommand{\yps}{\ensuremath{y_{\rm P1}}}

\newcommand{\grizy}{\gps\rps\ips\zps\yps}

\newcommand{\izy}{\ips\zps\yps}
\newcommand{\zy}{\zps\yps}

\newcommand{\iz}{\ips\zps}

\newcommand{\PS}{\protect \hbox {Pan-STARRS1}}

\slugcomment{Not to appear in Nonlearned J., 45.}

\shorttitle{First PS1 High Redshift Quasar}
\shortauthors{Morganson et al.}

\begin{document}


\title{The First High Redshift Quasar from Pan-STARRS}


\author{Eric Morganson\altaffilmark{1}, Gisella De Rosa\altaffilmark{1}, Roberto Decarli\altaffilmark{1}, Fabian Walter\altaffilmark{1}, Ken Chambers\altaffilmark{2}, Ian McGreer\altaffilmark{3},  Xiaohui Fan\altaffilmark{3}, William Burgett\altaffilmark{2}, Heather Flewelling\altaffilmark{2}, Klaus Hodapp\altaffilmark{2}, Nick Kaiser\altaffilmark{2}, Eugene Magnier\altaffilmark{2}, Paul Price\altaffilmark{4}, Hans-Walter Rix\altaffilmark{1}, Bill Sweeney\altaffilmark{2}, Christopher Waters\altaffilmark{2}}
\email{morganson@mpia.de}


\altaffiltext{1}{Max-Planck-Institut f\"ur Astronomie, K\"onigstuhl 17, 69117 Heidelberg, Germany}
\altaffiltext{2}{Institute for Astronomy, University of Hawaii at Manoa, Honolulu, HI 96822, USA}
\altaffiltext{3}{Steward Observatory, University of Arizona, 933 N Cherry Ave., Tucson, AZ, 85721, USA}
\altaffiltext{4}{Princeton University Observatory, 4 Ivy Lane, Peyton Hall, Princeton University, Princeton, NJ 08544, USA}


\begin{abstract} 
We present the discovery of the first high redshift ($z > 5.7$) quasar from the Panoramic Survey Telescope and Rapid Response System 1 (\PS\ or PS1). This quasar was initially detected as an $\ips$ dropoutout in PS1, confirmed photometrically with the SAO Widefield InfraRed Camera (SWIRC) at Arizona's Multiple Mirror Telescope (MMT) and the Gamma-Ray Burst Optical/Near-Infrared Detector (GROND) at the MPG 2.2 m telescope in La Silla. The quasar was verified spectroscopically with the the MMT Spectrograph, Red Channel and the Cassegrain Twin Spectrograph (TWIN) at the Calar Alto 3.5 m telescope. It has a redshift of 5.73, an AB $\zps$ magnitude of 19.4, a luminosity of 3.8$\times$10$^{47}$ erg s$^{-1}$ and a black hole mass of 6.9$\times$10$^9$ M$_{\odot}$. It is a Broad Absorption Line quasar with a prominent Ly-$\beta$ peak and a very blue continuum spectrum. This quasar is the first result from the PS1 high redshift quasar search that is projected to discover more than a hundred $\ips$ dropout quasars, and could potentially find more than 10 $\zps$ dropout ($z > 6.8$) quasars.  

\end{abstract}





\section{Introduction}\label{intro}
Quasars are massive black holes in the centers of galaxies that have large accretion rates and correspondingly large luminosities \citep{REES84,ANTO93,K&N99}. They can be up to one hundred times brighter than their host galaxies \citep[e.g.][]{VILL++06} and can thus be observed spectroscopically and analyzed in depth at higher redshifts than galaxies without quasars. High redshift ($z > 5.7$) quasars are an essential tool for probing the early universe. Obtaining a statistically complete set of $z \approx 6$ quasars constrains early structure evolution and black hole formation \citep[e.g.][]{JIAN++08}. Quasar spectra can be used to probe the evolution of metal abundances \citep{FREU++03,JIAN++07,KURK++07,KURK++09} . And by observing the Gunn-Peterson troughs \citep{G&P65} in the spectrum, one can put constraints on the neutral hydrogen fraction (H\Rmnum{1}) in the early universe \citep[e.g.][]{BECK++01,FAN++02,FAN06} and directly probe the end of cosmic reionization.  

Redshift $\approx 6$ quasars' usefulness as a probe of the early universe has led to significant interest and several extensive searches. \citet{FAN++01} first discovered them in the Sloan Digital Sky Survey \citep[SDSS,][]{YORK++00}. Several searches in the last decade have found a total of roughly 60 quasars at $z \approx 6$\citep{FAN++06,JIAN++08,WILL++10}. Extrapolating the statistics from \citet{JIAN++08} suggest that there are $\approx$ 450 redshift $6 < z < 7$, AB magnitude $z < 21$ quasars in the entire sky. 

Quasar searches have been slow to extend the redshift range of known quasars. \citet{FAN++01} included a $z = 6.28$ quasar and until 2011 subsequent searches only found quasars up to $z = 6.43$ \citep{FAN++03,WILL++07}. This is to be expected, as the main method of identifying $z \approx 6$ candidates is as "i dropouts" observable in the SDSS (or other surveys') $z'$ filter, but is either not observable or very faint in the $i'$ band. At high redshift, this large $i'-z'$ color is due to the Ly-$\alpha$ break at rest frame 1216 \AA producing a Gunn-Peterson trough. The transition between the SDSS $i'$ and $z'$ band is at $\approx$ 8200 \AA, and the $z'$ band transmission drops markedly at wavelengths greater than 9000 \AA\ \citep{FUJI++96}. This limits the observation of quasars to the $z < 6.4$ redshift range in SDSS and surveys with similar filter sets.

The UKIRT Infrared Deep Sky Survey \citep[UKIDSS,][]{LAWR++07} made using the United Kingdom Infrared Telescope (UKIRT) is capable of finding quasars at higher redshift than SDSS and similar surveys. The UKIDSS Large Area Survey (LAS) will eventually cover 7500 deg$^2$ of SDSS area in YJHK with average Y depth is 20.2. \citet{MORT++08} have begun searching for quasars up to $z = 7.2$ as $z'$ dropouts, sources detected in Y that are either faint or undetected in SDSS $z'$. UKIDSS's area and depth will allow it to detect a handful of $z = 7$ quasars, and \citet{MORT++11b} recently reported their first detection.  

In this paper, we discuss our $z \approx 6$ and $z \approx 7$ quasar search \citep{PRIC++07} in the Panoramic Survey Telescope and Rapid Response System 1 \citep[PS1][]{KAIS++02}. In the next section we discuss the PS1 dataset and why it is ideal for searching for both $z \approx 6$ and $z \approx 7$ quasars. In section \ref{sect:selection}, we explain how we currently select candidates, and how this selection can become more efficient as the project matures. In sections \ref{sect:followup1} and \ref{sect:followup2} we discuss the photometric and spectroscopic follow up that is necessary to confirm our quasar candidates. We present the first PS1 $z \approx 6$ quasar in section \ref{sect:quasar} and discuss its implications in section \ref{sect:conc}. 

\section{The PS1 Dataset}\label{sect:data}

PS1 \citep{KAIS++02,KAIS++10,CHAM11} is a 1.8 m optical telescope with a 7 degree field of view that can image the sky in the $\gps$, $\rps$, $\ips$ and $\zps$ filters which cover the $4000\rm{\AA} < \lambda < 9200\rm{\AA}$ spectral range similarly to the analogously-named SDSS $g'$, $r'$, $i'$ and $z'$ filters. It also has a $\yps$ filter which, including the spectral response of the camera, covers the $9200\rm{\AA} < \lambda < 10500 \rm{\AA}$ range. The telescope is producing several surveys including a solar system Near Earth Object survey, a Stellar Transit Survey, a Deep Survey of M31, a Medium Deep survey consisting of 11 PS1 footprints spaced around the sky and a $3\pi$ Survey which covers $3/4$ of the sky in all five bands \citep{CHAM11}. This final survey is the focus of our quasar search as its unique areal coverage provides the best opportunity to find $z \approx 6$ and $z \approx 7$ quasars. In the future, the PS1 Medium Deep survey will also allow us to search for fainter quasars, analogous to the faint quasar search in SDSS Stripe 82 in \citet{JIAN++08}

The PS1 $3\pi$ survey takes four exposures per year with each of the $\grizy$ filters. The yearly fill factor is roughly between 90\% and 95\% in each band. The missing area is due mostly to non-detection areas on the camera plane and weather restricting us to 2 or rarely 0 exposures in some areas of the sky. Individual $\grizy$ exposures have median 5$\sigma$ limiting AB magnitudes of 21.9, 21.8, 21.5, 20.7 and 19.7, respectively. These are median results of the last year of telescope performance, and recent hardware and software improvements have improved limiting magnitude by up to 0.3 magnitudes. Stacked images are not yet available, and the work presented here is based on single exposure detections. However, when stacks are made, we expect a single year's stacked image to increase each limiting magnitude by approximately 0.7 (accounting for some survey incompleteness), and the stacks of the proposed three year duration of the survey to increase them by 1.2. We see in Table \ref{tab:limmag} that in the $\iz$ bands, which are critical for detecting $z > 5.7$ quasars, PS1 will probe deeper than the SDSS. At 30000 deg$^2$ (minus approximately 4000 deg$^2$ since we do not search for quasars in the galactic plane), PS1 is also significantly larger than SDSS DR8 (14000 deg$^2$) and the UKIDSS LAS (7500 deg$^2$). Finally, the PS1 $\zps$ images are significantly redder than the SDSS $z'$ images, because the PS1 camera has a much higher efficiency in the $\lambda \approx 9000\rm{\AA}$ region. This should allow us to find redder $i$ dropout quasars than could be found with SDSS. 

\begin{table}
\begin{tabular}{cccccc}
	\hline
Filter &  SDSS    &   UKIDSS  & PS1 Exposure & PS1 1 Year & PS1 3 Year  \\
	\hline
u   & 22.3 & --   & --   & --   & --   \\
g   & 23.3 & --   & 21.9 & 22.6 & 23.1 \\
r   & 23.1 & --   & 21.8 & 22.5 & 23.0 \\
i   & 22.3 & --   & 21.5 & 22.2 & 22.7 \\
z   & 20.8 & --   & 20.7 & 21.4 & 21.9 \\
y/Y & --   & 21.1 & 19.7 & 20.4 & 20.9 \\
J   & --   & 20.9 & --   & --   & --   \\
H   & --   & 20.2 & --   & --   & --   \\
K   & --   & 20.2 & --   & --   & --   \\
	\hline
\end{tabular}
\caption{\rm{5$\sigma$ Limiting AB Magnitudes of SDSS\citep{YORK++00}, UKIDSS LAS \citep{LAWR++07,HEWE++06} and PS1 3$\pi$ (1 Year and 3 Year stack results are predicted). Similarly named filters from different surveys are not exactly the same.}}\label{tab:limmag}
\end{table}

This huge area and depth in the red bands will enable searches for many $z > 5.7$ quasars. \citet{FAN++06} found 19 $i'$ dropout quasars in 6600 deg$^2$. PS1 probes a half magnitude deeper in $\ips$ and covers nearly four times the searchable area. If we assume a quasar luminosity function of $\Phi(L) \propto L^{-3}$, we expect to find $\approx$ 200. Analogously, the PS1 area and $\zps$ depth will allow us to find several times as many quasars as the 7500 deg$^2$ UKIDSS LAS in which \citet{VENE07} expects to find approximately 10 $z >\ 6.5$. PS1 does not have JHK bands that UKIDSS uses to select quasars, so it is difficult to estimate exactly how much more efficient the $\zps$ depth will make the PS1 search.  

\section{Candidate Selection}\label{sect:selection}

PS1 is a young survey and neither its catalogs nor its images exist in their final form. Our current method of selecting candidates will change as the data and image processing is improved. We will thus describe how we currently find candidates and how we plan to do so when the PS1 stacked imaging and catalogs have been produced.

Currently, we require sources be detected twice in their detection band ($\yps$ for $\zps$ dropouts and $\zps$ for $\ips$ dropouts). The 3$\pi$ survey takes either 2 or 4 exposures of every part of the sky in each band every year, so this requirement does not reduce our effective survey area significantly. We require that these two detections not be flagged as cosmic rays, saturated pixels, defects, blended sources or other suspicious entities. We also require that the sources be either 1.5 magnitudes fainter or undetected in all bluer bands. In the future, we will raise this dropout requirement to 2 magnitudes to exclude brown dwarfs and other contaminants, but in the first run we were interested in characterizing our false detections. We require that candidates be fainter than AB magnitude 18.5 in their detection band (the $z'$ magnitude of the brightest SDSS $i'$ dropout). This eliminates some extra spurious sources near bright galaxies and stars. For $\ips$ dropouts, we require that they be brighter than $\zps = 20.5$, since a huge number of faint $\ips - \zps < 1.5$ objects will be undetected in bluer bands. We also require that objects be more than 10 degrees away from the galactic equator to avoid confusion by galactic stars and most galactic reddening.

When we performed this search, the PS1 database did not differentiate between sources that are covered by good imaging in a band but not detected in that band and sources that are not covered by good imaging (or any imaging). The vast majority of apparent dropout were simply not imaged in bluer bands. We eliminated false candidates by requiring that they be within the circular approximation of at least one exposure in every filter. We also require a dropout band detection (even a defect or cosmic ray) within 5" of our source to ensure that there is data coverage in the immediate area of the detection. When the data is ultimately stacked, we will have a magnitude or limiting magnitude in every filter at every point and this step will be unnecessary. Finally, we made sure that each candidate has no neighbor (excluding defects and cosmic rays) within 2" in any band bluer that was less than 0.5 magnitudes dimmer than the candidates in the detection band. This last requirement eliminated sources whose blue detections were not matched to their red detections by the PS1 software. This left $10^5$ sources, a small enough number to study at the pixel level.

At the time of this writing, PS1 has just begun producing 3$\pi$ stacked images and corresponding catalogs which assign a limiting magnitude to sources that are not detected in a given filter and indicate which sources are only imaged in some bands. In addition, a second year of data is already filling in the areas which were not covered due to chip gaps and other holes in our images. So the number of false positives at this stage is decreasing exponentially. 

We obtained postage stamps of the $10^5$ candidates which passed catalog level filtering from the Postage Stamp Server maintained by the PS1 Image Processing Pipeline team at the Institute for Astronomy. For each candidate, we obtained 24" postage stamps of every exposure from the dropout band and detection band. We produce weighted mean stacks of these exposures and performed forced photometry on each stacked image. Approximately half of all candidates were in small image gaps in the detection band. We automatically eliminated all sources which were not detected in the detection band with less than 0.15 magnitude uncertainty in our postage stamp stacks. This cut was judged, by eye, to be complete for real sources. Approximately half of the candidates were saturation defects not spotted by the PS1 pipeline. We required that sources be 1.5 magnitudes fainter in the dropout band than in the detection band if they are detected or have a 5$\sigma$ limiting magnitude 1.5 magnitudes greater than the detection magnitude. We also required that sources have 80\% of their PSF flux imaged by good pixels to ensure that they are not chip edge or other artifacts. In cases where an object had been imaged in the detection band on two nights, we eliminated sources that were detected one night but not on another night of sufficient depth.

Finally, we examined the stacks and individual exposures of the remaining 3$\times$10$^3$ sources by eye. At this level, we eliminated cosmic rays that the software missed, marginal detections and artifacts registered as sources. There were approximately 10$^3$ candidates over the half of the 3$\pi$ we examined. We prioritized our follow up candidates based on the reliability of the detection, the size of the color difference and observability from the observation site. So far, we have performed some follow up on $\approx 200$ quasars. 

\section{Candidate Follow Up: Photometry}\label{sect:followup1}

We require more information than PS1 gives to confirm a candidate. In individual exposures, most $z > 5.7$ quasars will be near the PS1 detection limit, and it is difficult to reject asteroids which are detected in the detection band but have moved when that area is covered by other bands, so simply confirming the existence of a candidate is important. In addition, brown dwarfs and other red objects can have similar PS1 colors to high redshift quasars \citep{FAN99}, so obtaining deeper $izy$ photometry for $\ips$ dropouts and $zyJ$ photometry for $\zps$ dropouts is essential to remove false detections. We confirm our candidates with photometry and spectroscopy from small telescopes before moving on to the more expensive spectroscopy necessary for deeper analysis. 

Initially, we planned to confirm candidates using the Calar Alto 3.5 m telescope and the Omega 2000 camera with the Y and J filters in spring 2011. Unfortunately, the telescope was offline due to a mechanical failure for a period of many months that included our observation time. We obtained five nights of time, 2011 February 16-20, with MMT SWIRC. For each source, we took nine 30 second exposures in Y band. This initial sample of sources was chosen to aggressively probe very faint sources, allowing for some false detections as a result of statistical noise. In addition, we did not account for moving solar system objects in our search. Only 23 of the 100 sources we followed up on were confirmed. This was to some degree expected, as coincident noise peaks and slow-moving objects which can be detected twice on one night are both plentiful in our large dataset. This work helped us understand how to select sources efficiently near our limiting magnitude.

\begin{figure}[ht]
\includegraphics[width=0.98\columnwidth]{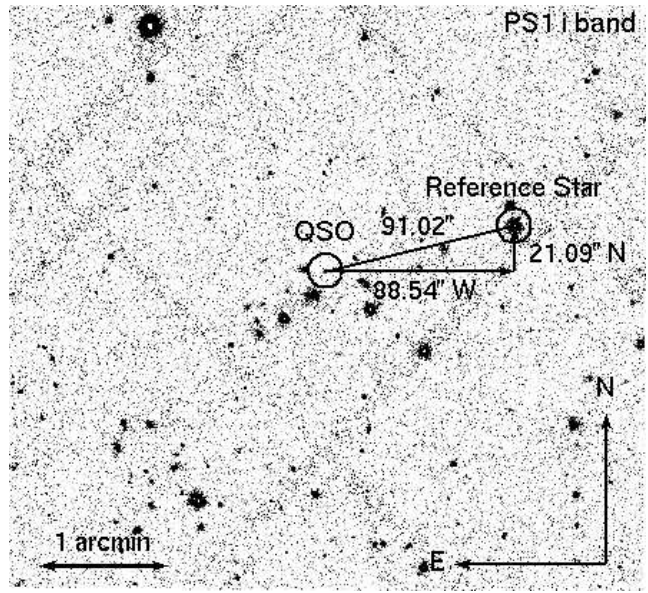}
\caption{The finding $\ips$ band finding chart of PSO J215.1512-16.0417 (located at J2000 215.1512, -16.0417 or 14$^h$20$^m$36.3$^s$, -16$^\circ$02$^m$30.2$^s$) with a guide star at 215.1258, -16.0358 or 14$^h$20$^m$30.2$^s$, -16$^\circ$02$^m$09.1$^s$.}
\label{fig:chart}
\end{figure}

\begin{figure}[ht]
\includegraphics[width=0.49\columnwidth]{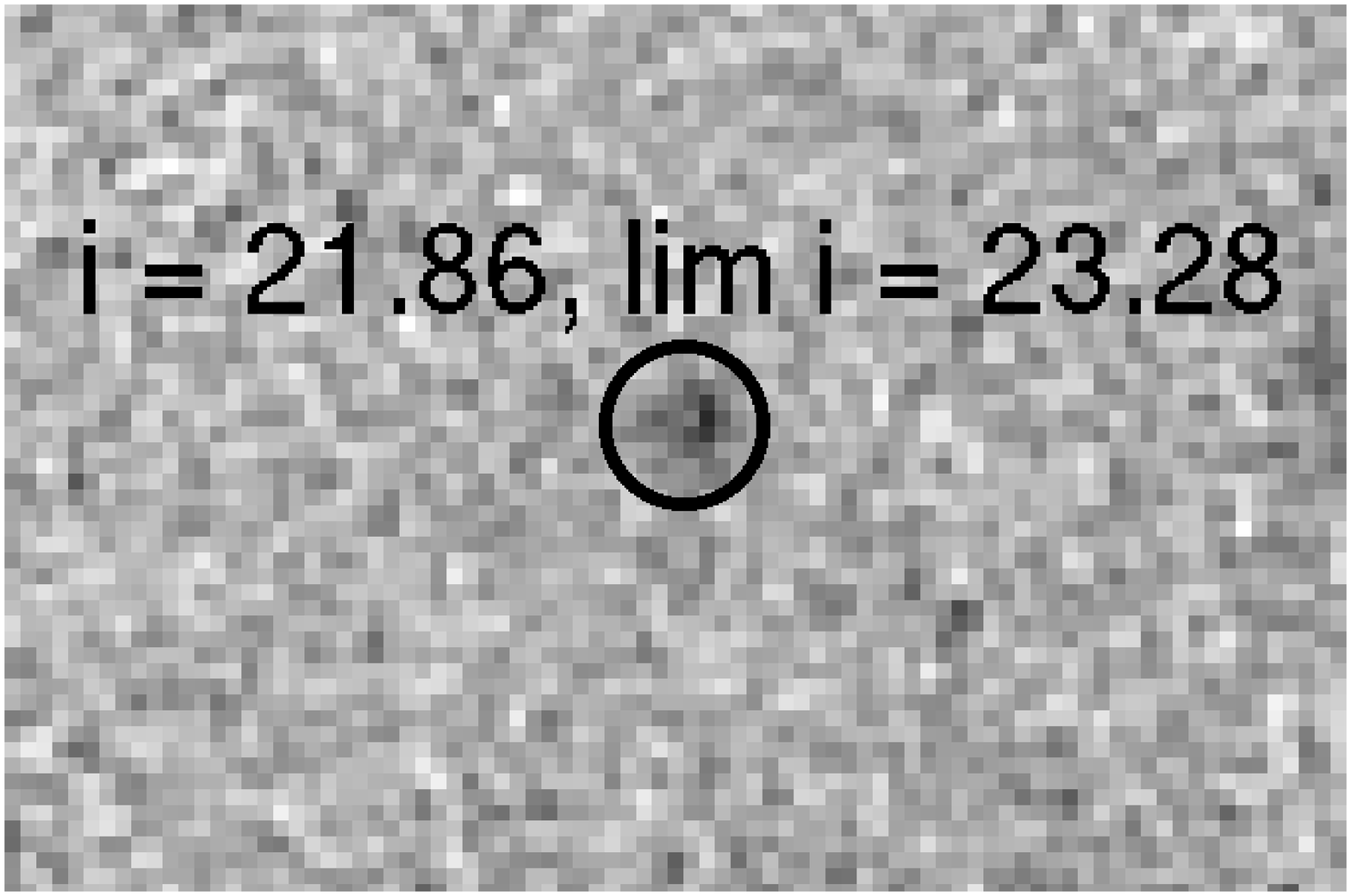}
\includegraphics[width=0.49\columnwidth]{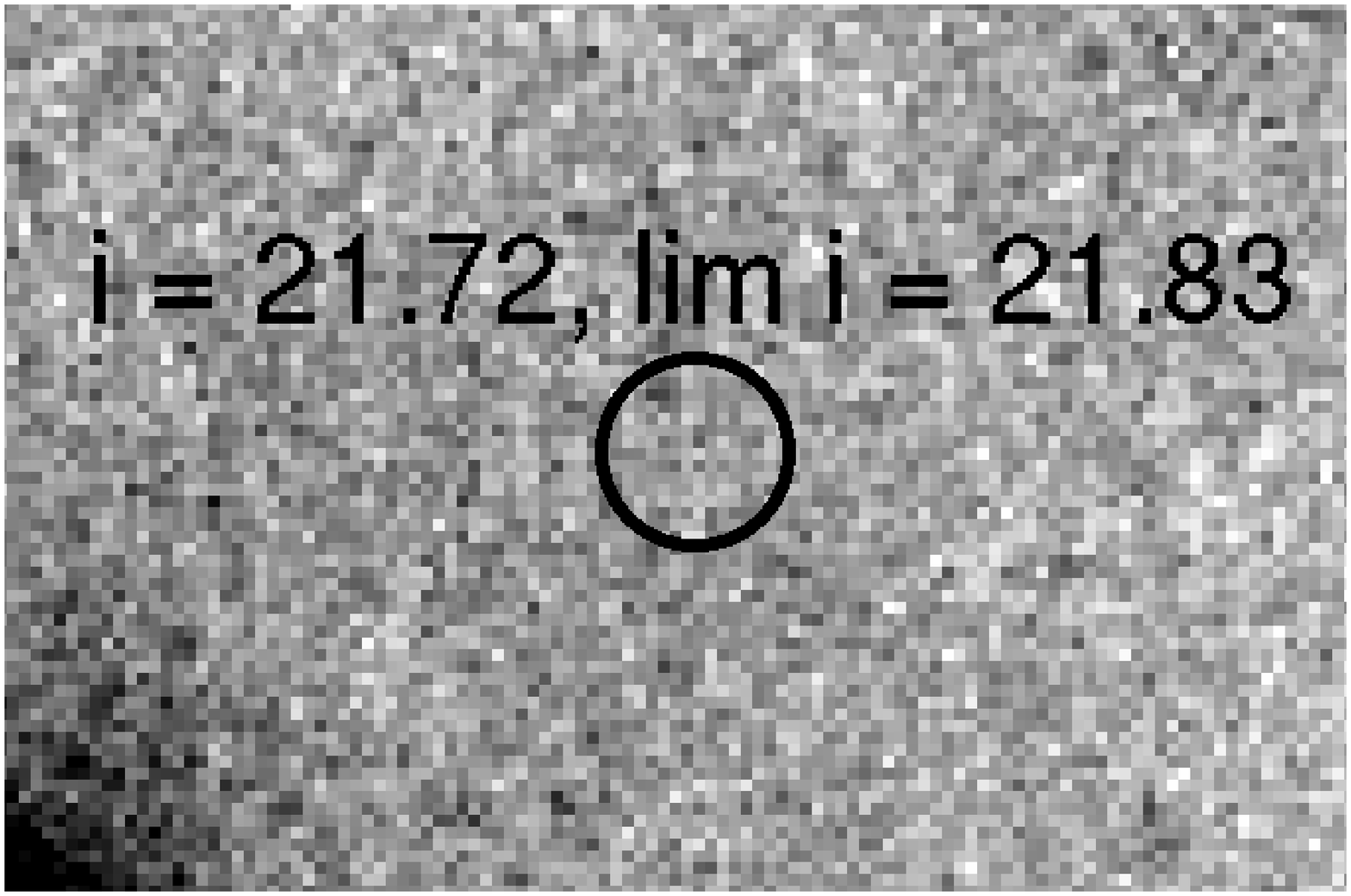}
\includegraphics[width=0.49\columnwidth]{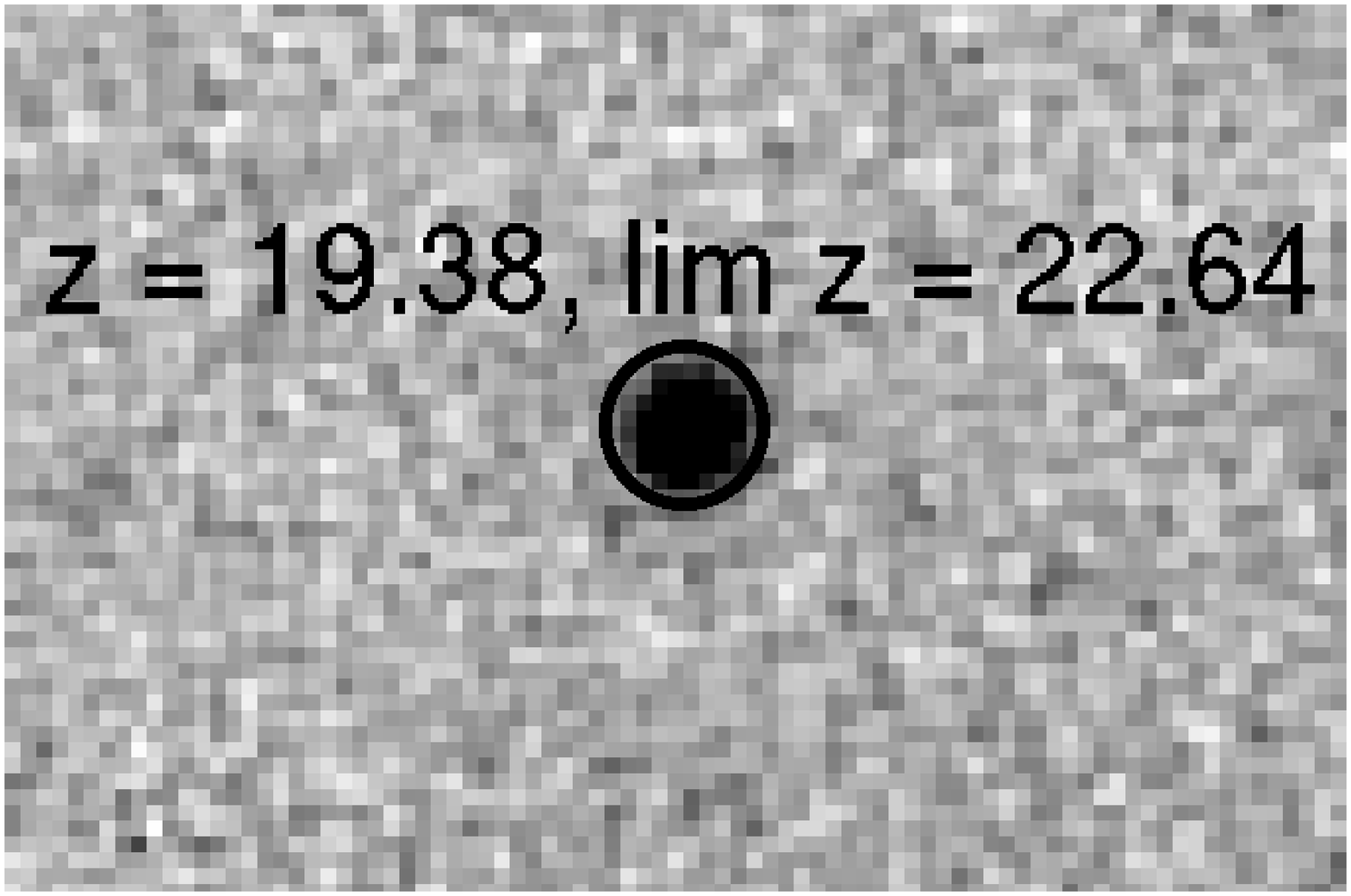}
\includegraphics[width=0.49\columnwidth]{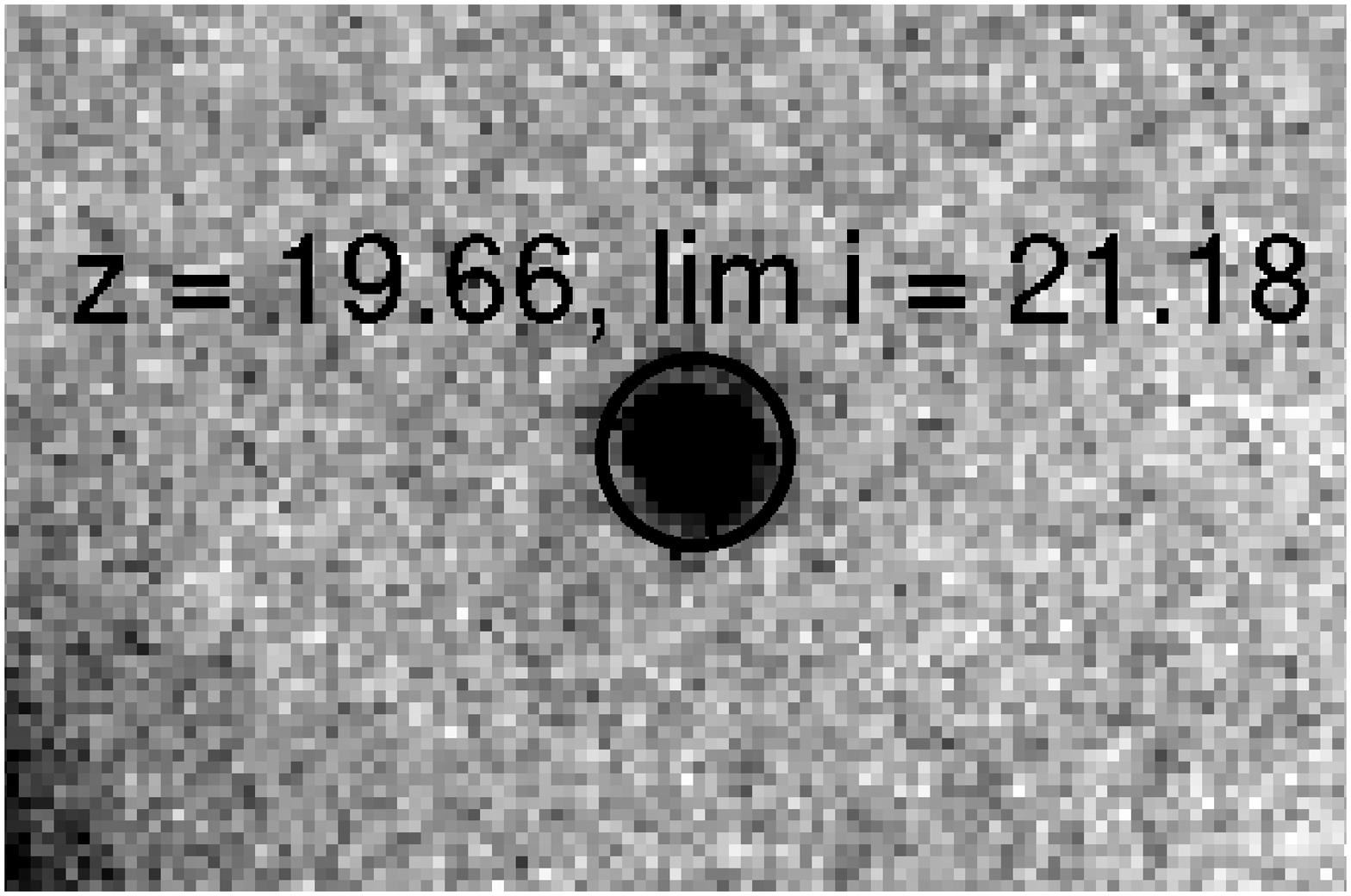}
\includegraphics[width=0.49\columnwidth]{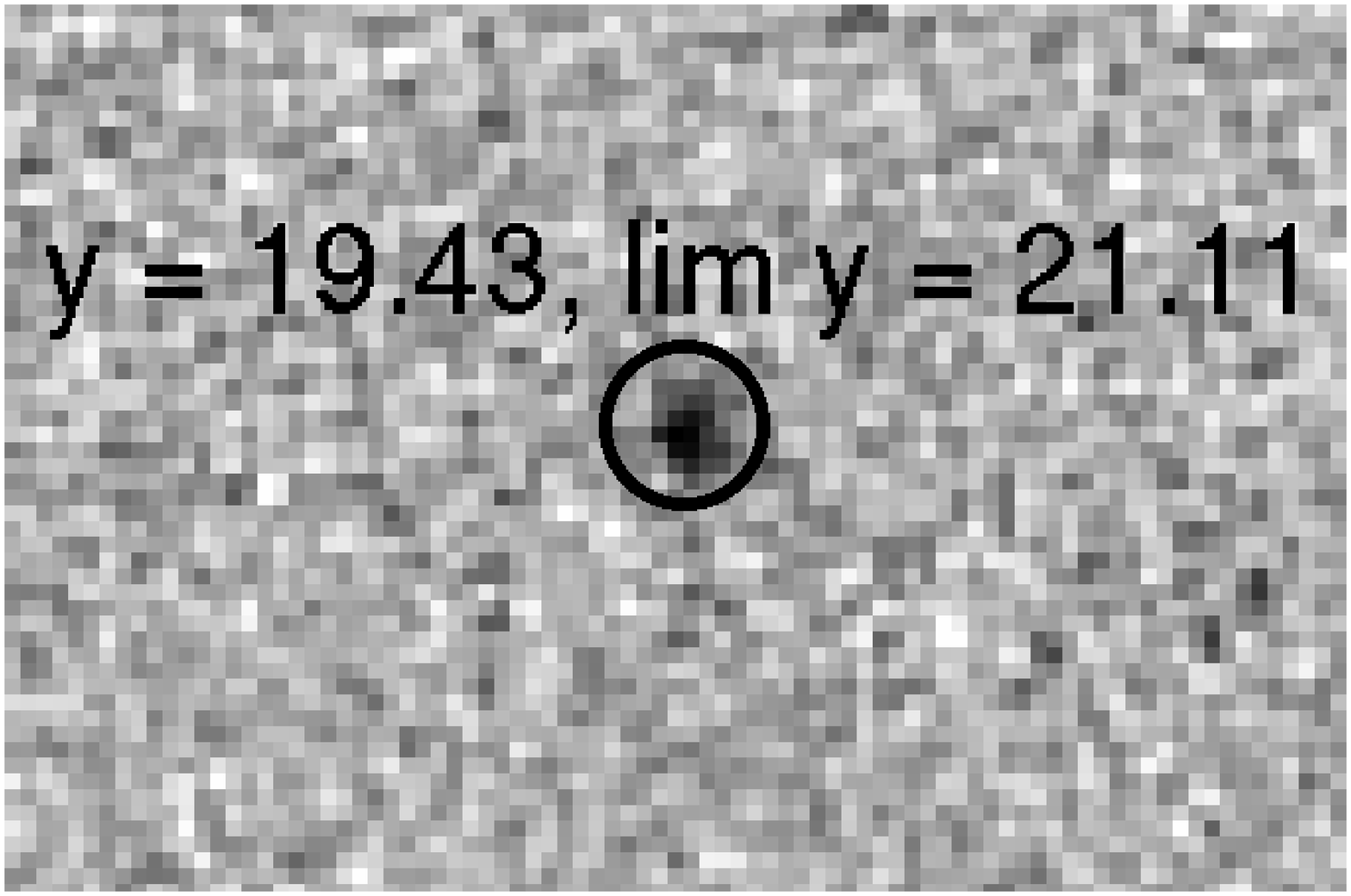}
\includegraphics[width=0.49\columnwidth]{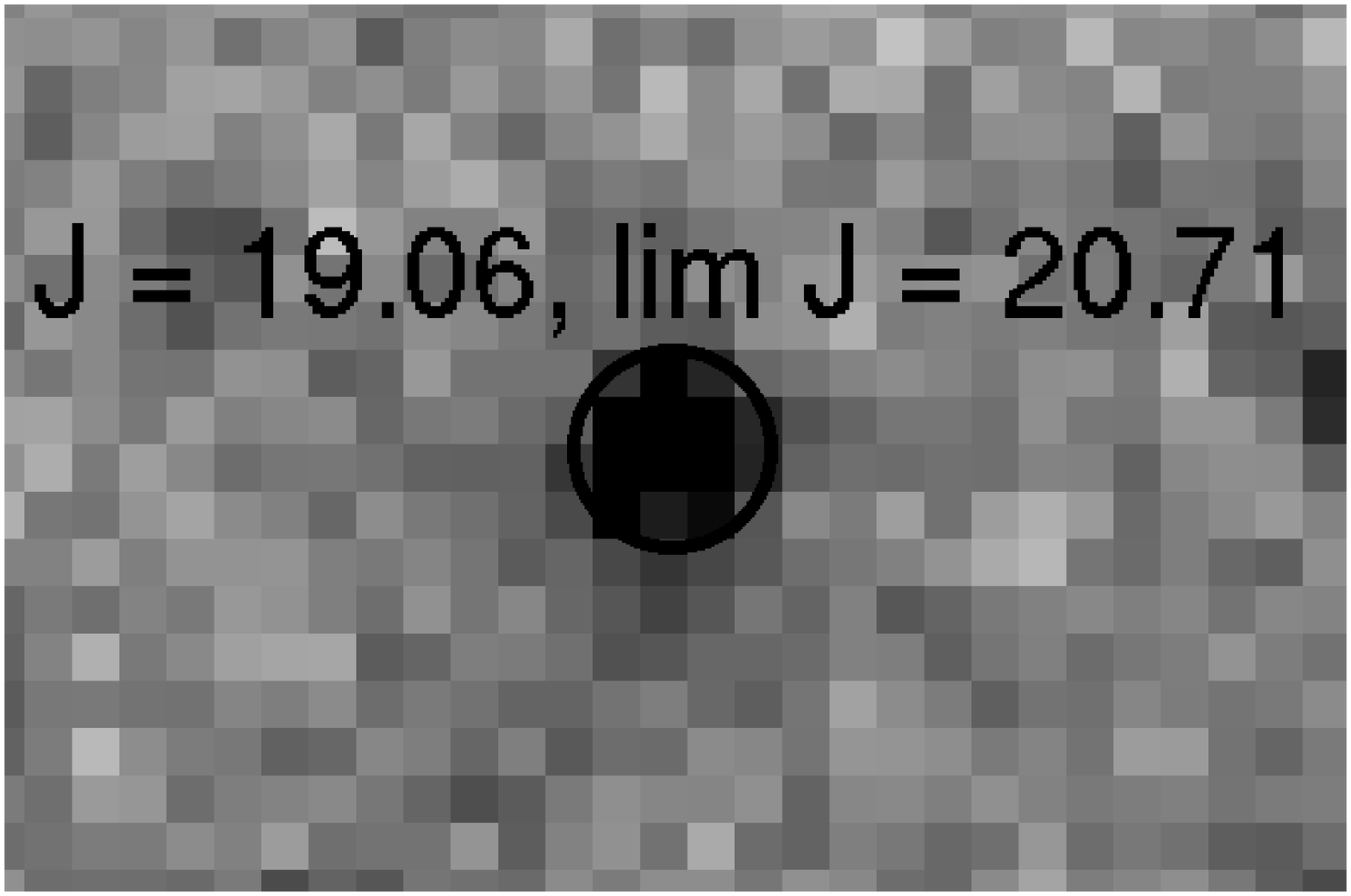}
\caption{The stacked postage stamps of  PSO J215.1512-16.0417 (located at J2000 215.1512, -16.0417 or 14$^h$20$^m$36.3$^s$, -16$^\circ$02$^m$30.2$^s$) in $\ips$ (top) $\zps$ (middle) and $\yps$ (bottom) on the left and the follow up images in $i_{GROND}$ (top) $z_{GROND}$ (middle) and $J_{GROND}$ (bottom) on the right. Each stamp includes AB magnitude and 5$\sigma$ limiting magnitude.}
\label{fig:GROND}
\end{figure}

We used GROND, a $grizJHK$ simultaneous imager at the 2.2 m telescope in La Silla, to obtain colors of the confirmed candidates from our MMT run and 61 additional candidates. The observations occupied roughly 40\% of a ten night observation period, 2011 March 4-13. Each object was observed using am instrument standard "8 minute observation block" in which the infrared images get a total of eight minutes exposure time in the optical bands get slightly more \citep{GREI++08}. We detected 48 of 62 sources, but found that most of them were less red than they had appeared in PS1. This was expected, since our typical PS1 colors have uncertainties of a few tenths, and we find many common bluer objects which randomly appear red in our PS1 observation. We did find one very promising candidate (finding chart in Fig. \ref{fig:chart} and multi-filter imaging in Fig. \ref{fig:GROND}), PSO J215.1512-16.0417 (at J2000 215.1512, -16.0417 or 14$^h$20$^m$36.3$^s$, -16$^\circ$02$^m$30.2$^s$). It had $i_{\rm{GROND}}$-$z_{\rm{GROND}}$ = 2.06 and $z_{\rm{GROND}}$-$J_{\rm{GROND}}$ = 0.60 (complete photometry in Table \ref{tab:GROND}). Unfortunately, this source was located near a chip gap in the GROND K detector, preventing us from reducing the K band image and obtaining accurate K photometry. Using individual K exposures, we were only able to set a limiting magnitude of approximately K > 17 for this source. 

\begin{table}
\begin{tabular}{cc}
        \hline
Filter & AB Magnitude\\
        \hline
$g_{GROND}$         &  < 22.51\\
$r_{GROND}$         &  < 22.57\\
$i_{GROND}$         & 21.72 $\pm$ 0.17  \\
$z_{GROND}$         & 19.66 $\pm$ 0.05  \\
$Y_{SWIRC}$         & 18.77 $\pm$ 0.06 \\
$J_{GROND}$         & 19.06 $\pm$ 0.08  \\
$H_{GROND}$         & 19.34 $\pm$ 0.14  \\
        \hline
\end{tabular}
\caption{\rm{GROND and SWIRC AB photometry of PSO J215.1512-16.0417 with statistical error bars or 5$\sigma$ limiting magnitudes (where applicable). The $y$ magnitude is SWIRC. All others are GROND.}}\label{tab:GROND}
\end{table}

We reduced the data using IRAF\footnote{IRAF is distributed by the
National Optical Astronomy Observatories, which are operated by the
Association of Universities for Research in Astronomy, Inc., under
cooperative
agreement with the National Science Foundation.} routines. Images were
processed with pipelines developed for this purpose by our group. Individual
frames were bias and dark subtracted and flat-fielded. Median sky images,
created by stacking individual exposures without realignment, were scaled
to the median counts in each frame and subtracted. Images were then
re-aligned and combined. The astrometric solution was computed using the
Astrometry.net software \citep{LANG++10}. Photometric zero points were
computed by comparing instrumental magnitudes of field stars with the
values reported in the 2MASS \citep{SKRU++06} and PS1 databases.
Typical photometric uncertainties were $\sim0.05$ mag. 

\section{Candidate Follow Up: Spectroscopy}\label{sect:followup2}

We had spectroscopic follow up time with both the MMT Red Channel Spectrograph (2011 March 12-13) and the Calar Alto TWINS (2011 April 9-12). We used a 1" slit at MMT and a 1.5" slit at Calar Alto. We examined PSO J215.1512-16.0417 and 13 others which had a less dramatic dropout color and were thus less likely to be quasars. The majority of these candidates were $\ips$ dropouts which we had not examined with GROND. We confirmed that our most likely candidate was indeed a quasar (as we discuss in Section \ref{sect:quasar}). We tentatively identified the other candidates as M, L and T dwarfs as well as three galaxies with particularly bright emission lines in our observation band, and have performed no additional follow up on them. 

Finally, we used the LBT NIR Spectroscopic Utility with Camera and Integral-Field Unit for Extragalactic Research (LUCI) at the Large Binocular Telescope (LBT) to measure the near IR spectrum of our quasar and estimate the continuum power law on 2011 May 28. We used a 1" slit. 

Optical spectroscopic data were reduced using standard IRAF tools. After bias
subtraction and flat-field correction, we aligned exposures by matching
the position on the slit of bright field sources. Frames were combined using
the \textsf{crreject} algorithm to get rid of cosmic rays. Wavelength
calibration was achieved through the acquisition of He/Ar arc spectra,
and compared with the observed sky emission lines (mainly OH) to correct
for instrument flexure. Typical uncertainties in the wavelength
calibration are $\sim 0.2$ \AA{} over the observed range. Spectra of
spectrophotometric standard stars were collected, in order to perform
relative flux calibration. 

LUCI observations were collected using the
standard nodding technique for NIR spectroscopy. Data were reduced using
our own IRAF-based pipeline. Differences between frames in A and B
position were computed in order to achieve optimal sky subtraction. We
aligned and stacked the A-B and B-A frames independently, applied
wavelength calibration then extracted and combined the resulting 1D
spectra. Dispersion solution is computed from the OH lines observed on
the target spectra. Correction for telluric absorption and flux
calibration were achieved by comparing the spectrum of the telluric
standard Hip71451 with the model of the model spectrum of a B9
star\footnote{http://www.eso.org/sci/facilities/paranal/instruments/isaac/tools/lib/index.html}.
Absolute flux calibration for both optical and NIR spectra was performed
by scaling them to match the GROND photometry in $z$ or $J$. GROND was in turn calibrated to 2MASS.



\section{The First PS1 High Redshift Quasar}\label{sect:quasar}

\begin{figure}[ht]
\includegraphics[width=0.99\columnwidth]{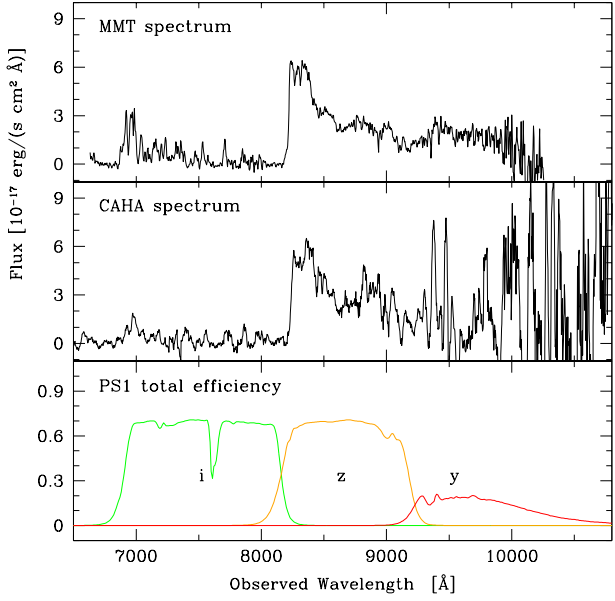}
\caption{The spectra of quasar PSO J215.1512-16.0417 obtained by MMT (top) and Calar Alto (Middle) as well as the PS1 $\izy$ filter curves (bottom).}
\label{fig:spectrum}\end{figure}

\begin{figure}[ht]
\includegraphics[width=0.99\columnwidth]{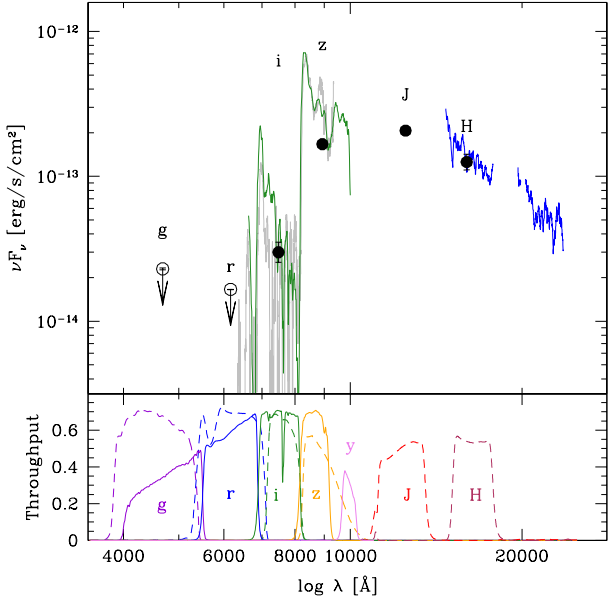}
\caption{The observer frame MMT (green) and LBT spectra (blue) of PSO J215.1512-16.0417 with GROND photometry measurements indicated (top). The bottom panel shows the PS1 and six observed GROND filter curves.}
\label{fig:spectrum2}\end{figure}

PSO J215.1512-16.0417, the candidate we identified as our most likely quasar (and not coincidentally the first candidate we observed spectroscopically) was confirmed as a quasar at MMT and Calar Alto. We present the spectrum from each observatory and the PS1 filter curves in Fig. \ref {fig:spectrum}. Comparing the spectra to the filter curves, we see that if this object were at significantly lower redshift, it would have had significant $\ips$ flux and would not have been detected as an $\ips$ dropout. In Fig. \ref{fig:spectrum2} we include the LBT NIR spectrum and compare it to the GROND photometry. 

\begin{figure}[ht]
\includegraphics[width=0.99\columnwidth]{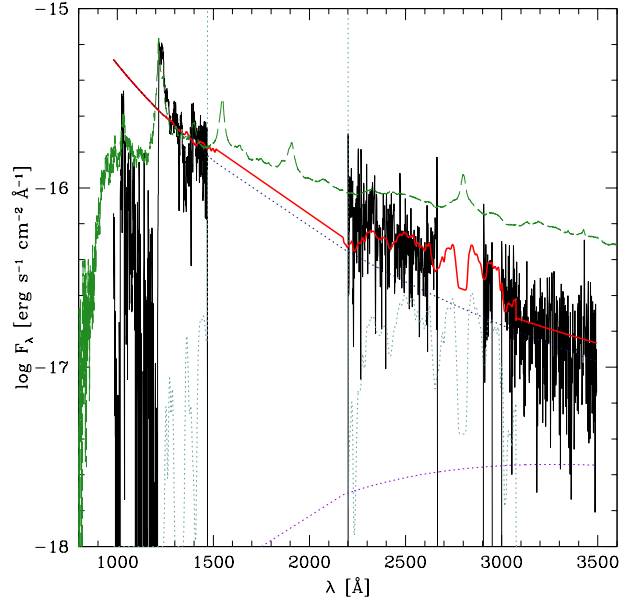}
\caption{The rest frame MMT and LBT spectra of PSO J215.1512-16.0417 (black), continuum-Balmer-iron model (red, see \citet{KURK++07} for details) and continuum fit (dotted). We compare it to the SDSS average high redshift quasar spectrum (green) from \citet{VAND++01}. }
\label{fig:lbt}\end{figure}

\begin{figure}[ht]
\includegraphics[width=0.99\columnwidth]{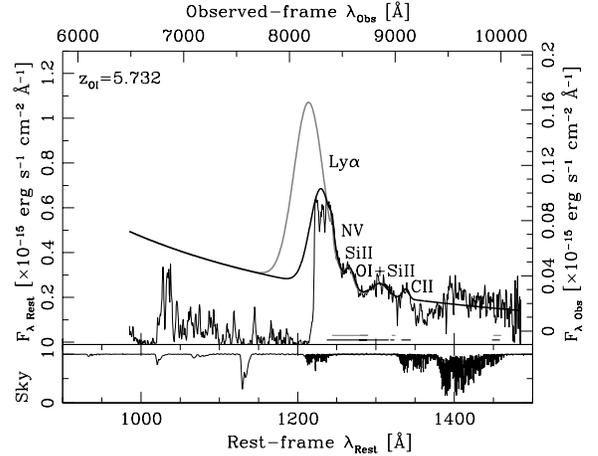}
\caption{The rest frame MMT spectrum of PSO J215.1512-16.0417 fitted with pertinent emission lines (excluding Ly-$\beta$) and continuum. We present the BAL model in which the Ly-$\alpha$ wavelength is fixed (gray) and the model in which the Ly-$\alpha$ wavelength is free (black).}
\label{fig:fit}\end{figure}

We analyze this quasar in figures \ref{fig:lbt} and \ref{fig:fit}. We use the long infrared/optical baseline that our MMT and LBT spectra provide in Fig. \ref{fig:lbt} to find that our continuum is well fit by a $\lambda^{-3.05}$ power law, much bluer than the SDSS $z \approx 6$ average of 1.3 \citep{VAND++01}. This estimate was made using the non-emission region around 1420 \AA\ from the MMT spectrum and the entire LBT spectrum following \citet{DERO++11} (the exponent has statistical error of 0.05) 

Having modeled the continuum, we fit the Ly-$\alpha$, N {\sc v}, Si {\sc ii}, O {\sc i}, and C {\sc ii} lines as Gaussians, masking high absorption regions where appropriate. We fit Ly-$\alpha$ N {\sc v} Si {\sc ii} simultaneously because they separated by less than a FWHM. But the other peaks are fitted independently. See Table \ref{tab:fits} for details. We do not fit the Si {\sc iv}-O {\sc iv} complex at 1400 \AA\ because of broad absorption features. The Ly-$\alpha$ peak is only a marginal detection and is several \AA\ from its expected location. This and a small Si {\sc iv} absorption at observer 9250 \AA\ (see Fig. \ref {fig:spectrum}) leads us to believe that this quasar is a Broad Absorption Line (BAL) quasar. We thus refit the Ly-$\alpha$-N {\sc v}-Si {\sc ii}, fixing the Ly-$\alpha$ peak using the O {\sc i} redshift. We present both the "free" (non-BAL) and "fixed" (BAL) results. For the remainder of this paper, we use values consistent with the BAL assumption. 

We estimate the redshift to be 5.7321 $\pm$ 0.0065 using the O {\sc i} line in Fig. \ref{fig:fit}, which is our most statistically robust emission line. Using the "fixed" fit, the lines have a measured redshift scatter of 0.02 (see Table \ref{tab:fits}), which represents the astrophysical systematic uncertainty. If we refit our peaks with the (SDSS average) $\lambda^{-1.3}$ power law continuum, the redshifts only shift by 0.004, so our dominant source of uncertainty appears to be astrophysical. Our final redshift estimate is 5.73 $\pm$ 0.02.

\begin{table*}
\begin{tabular}{cccccc}
	\hline
Line &  $\lambda_{\rm peak}$ &  $z$  & Flux & FWHM & EW\\
 &  \AA  &     & $10^{-15}\rm{erg}\rm{cm}^{-2}\rm{s}^{-1}$ & \AA & \AA\\
	\hline
Ly-$\alpha$ (free)  & 1231    $\pm$ 18   & 5.810  $\pm$ 0.098  & 15   $\pm$ 12   & 33   $\pm$ 20  & 107 $\pm$ 83  \\
N {\sc v} (free)           & 1244.97 $\pm$ 0.89 & 5.7465 $\pm$ 0.0048 & 0.38 $\pm$ 0.43 & 6.5  $\pm$ 4.0 & 2.5 $\pm$ 2.9  \\
Si {\sc ii} (free)         & 1268.49 $\pm$ 0.97 & 5.7554 $\pm$ 0.0051 & 1.26 $\pm$ 0.32 & 13.1 $\pm$ 2.0 & 8.0 $\pm$ 2.0  \\
Ly-$\alpha$ (fixed) & 1215.67            & 5.7321              & 38   $\pm$ 17   & 44.4 $\pm$ 5.5 & 280 $\pm$ 130  \\
N {\sc v} (fixed)          & 1245.36 $\pm$ 0.39 & 5.7567 $\pm$ 0.0022 & 0.23 $\pm$ 0.15 & 2.9  $\pm$ 1.4 & 1.5 $\pm$ 1.0  \\
Si {\sc ii} (fixed)        & 1267.21 $\pm$ 0.59 & 5.7567 $\pm$ 0.0032 & 1.14 $\pm$ 0.20 & 12.7 $\pm$ 1.6 & 7.3 $\pm$ 1.3  \\
O {\sc i}          & 1306.9  $\pm$ 1.3  & 5.7321 $\pm$ 0.0065 & 1.24 $\pm$ 0.25 & 22.1 $\pm$ 3.7 & 7.4 $\pm$ 1.4  \\
C {\sc ii}         & 1339.50 $\pm$ 0.72 & 5.7452 $\pm$ 0.0036 & 0.43 $\pm$ 0.11 & 9.4  $\pm$ 2.1 & 2.35 $\pm$ 0.64  \\
	\hline
\end{tabular}
\caption{\rm{Emission Line Peaks from PSO J215.1512-16.0417. All error bars are statistical. We include the fitted central wavelength, integrated flux, line width, FWHM, and equivalent width. Because of the (likely) broad absorption feature, we fit the Ly-$\alpha$-N {\sc v}-Si {\sc ii} complex with a free Ly-$\alpha$ peak and a Ly-$\alpha$ with a peak fixed by the O {\sc i} redshift measurement.}}\label{tab:fits}
\end{table*}

We have no Mg {\sc ii} or C {\sc iv} line in our spectrum and cannot measure our quasar's black hole mass and bolometric luminosity in the most precise way. Instead, we measure the flux at 1350 \AA\ is 1.89$\times$10$^{-16}$ erg \AA$^{-1}$ cm$^{-2}$ s$^{-1}$. Multiplying by $4 \pi D_L^2 \lambda$ yields a monochromatic luminosity of 9.5$\times$10$^{46}$ erg s$^{-1}$. We multiply this by the bolometric coefficient of 3.81 from \citet{RICH++06}, to obtain a bolometric luminosity of 3.8$\times$10$^{47}$ erg s$^{-1}$ with 0.2 dex of uncertainty. This in turn corresponds to an Eddington Mass of 2.9$\times$10$^9$ M$_{\odot}$. In an analysis of the complete set of SDSS high redshift quasars, \citet{DERO++11} find that bright ($\rm{L} > 10^{47}$) $z \approx 6$ quasars have typical Eddington ratios of 0.43 with 0.2 dex dispersion. We thus estimate our black hole mass to be 6.9$\times$10$^9$ M$_{\odot}$ with 0.3 dex uncertainty. 

\begin{figure}[ht]
\includegraphics[width=0.99\columnwidth]{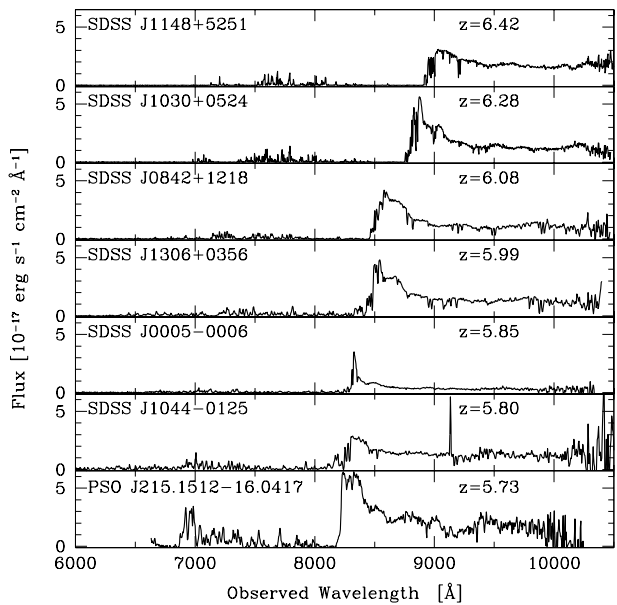}
\caption{Comparison of PSO J215.1512-16.0417 to SDSS $z \approx 6$ quasars.}
\label{fig:compare}\end{figure}

When we compare PSO J215.1512-16.0417 to other $z \approx 6$ quasars (Fig. \ref{fig:compare}), we see that it has at least three interesting features. First, it appears to be BAL quasar which some studies have found is true of only 10\% of all quasars \citep{TRUM++06}. In addition, PSO J215.1512-16.0417 has a very bright Ly-$\beta$ peak that will allow us to more precise study the hydrogen absorption along the line of sight in the future. PSO J215.1512-16.0417 is also one of the bluest known at this redshift. A more complete analysis and comparison of quasars will be described in \citet{DERO++11}. 

\section{Discussion and Conclusions}\label{sect:conc}

PSO J215.1512-16.0417 is a very blue, (likely) BAL quasar with a prominent Ly-$\beta$ peak that demonstrates the ability of PS1 to find $z \approx 6$ quasars. We expect to find 100s of $\ips$ dropouts that will allow us to make a statistically precise measurement of the $z \approx 6$ universe. The survey also offers the exciting prospect of extending our measurements of Gunn-Peterson troughs, emission and absorption lines, and black holes masses to the $z = 7$ regime.   

Continued development of PS1 survey, hardware and software as well as our experience regarding follow up observations will greatly increase our efficiency. At the time of our initial candidate selection, the $\iz$ and $\zy$ overlap regions were only a few thousand square degrees. In addition, several improvements to the PS1 telescope have decreased the background levels and FWHMs, increasing the limiting magnitude by significantly. As PS1 finishes its first year, the data processing software has been (tentatively) finalized and many false positives in the database have been eliminated. This will allow us to examine more candidates. In addition, PS1 has recently begun producing stacked data which should add 0.7 magnitudes to the limiting magnitude and allow us to discriminate between sources which were not imaged in a band and sources which were imaged but too faint to be detected. Finally, our follow up observations have emphasized the necessity of obtaining deep $izy$ observations for $\ips$ dropouts and deep $zyJ$ observations for $\zps$ dropouts.   

PSO J215.1512-16.0417 is our first glimpse into a bright future $z > 5.7$ discovery with PS1. With this survey we expect to vastly extend the statistical power and redshift range of high redshift quasar studies. The fact that we were able to find an interesting quasar so early in the project, while the survey was incomplete, processing software was incomplete, hardware had not yet been fine-tuned and stacked images had not been made demonstrates how powerful a tool PS1 will be for finding quasars in the early universe.

\section{Acknowledgments}

The PS1 Surveys have been made possible through contributions of the Institute for Astronomy, the University of Hawaii, the Pan-STARRS Project Office, the Max-Planck Society, and its participating institutes, the Max Planck Institute for Astronomy, Heidelberg, and the Max Planck Institute for Extraterrestrial Physics, Garching, The Johns Hopkins University, Durham University, the University of Edinburgh, Queen's University Belfast, the Harvard-Smithsonian Center for Astrophysics, and the Las Cumbres Observatory Global Telescope Network, Incorporated, the National Central University of Taiwan, and the National Aeronautics and Space Administration under Grant No. NNX08AR22G issued through the Planetary Science Division of the NASA Science Mission Directorate.

GDR is supported by the Deutsche Forschungsgemeinschaft priority program
1177 (''Witnesses of Cosmic History: Formation and evolution of galaxies,
black holes and their environment'').

RD acknowledges funding from Germany's national research center for aeronautics and space (DLR, project FKZ 50 OR 1104).



\bibliography{ms}

\begin{thebibliography}{37}
\expandafter\ifx\csname natexlab\endcsname\relax\def\natexlab#1{#1}\fi

\bibitem[{{Antonucci}(1993)}]{ANTO93}
{Antonucci}, R. 1993, \araa, 31, 473

\bibitem[{{Becker} {et~al.}(2001){Becker}, {Fan}, {White}, {Strauss},
  {Narayanan}, {Lupton}, {Gunn}, {Annis}, {Bahcall}, {Brinkmann}, {Connolly},
  {Csabai}, {Czarapata}, {Doi}, {Heckman}, {Hennessy}, {Ivezi{\'c}}, {Knapp},
  {Lamb}, {McKay}, {Munn}, {Nash}, {Nichol}, {Pier}, {Richards}, {Schneider},
  {Stoughton}, {Szalay}, {Thakar}, \& {York}}]{BECK++01}
{Becker}, R.~H., {Fan}, X., {White}, R.~L., {Strauss}, M.~A., {Narayanan},
  V.~K., {Lupton}, R.~H., {Gunn}, J.~E., {Annis}, J., {Bahcall}, N.~A.,
  {Brinkmann}, J., {Connolly}, A.~J., {Csabai}, I., {Czarapata}, P.~C., {Doi},
  M., {Heckman}, T.~M., {Hennessy}, G.~S., {Ivezi{\'c}}, {\v Z}., {Knapp},
  G.~R., {Lamb}, D.~Q., {McKay}, T.~A., {Munn}, J.~A., {Nash}, T., {Nichol},
  R., {Pier}, J.~R., {Richards}, G.~T., {Schneider}, D.~P., {Stoughton}, C.,
  {Szalay}, A.~S., {Thakar}, A.~R., \& {York}, D.~G. 2001, \aj, 122, 2850

\bibitem[{{Chambers}(2011)}]{CHAM11}
{Chambers}, K.~C. 2011, in Bulletin of the American Astronomical Society,
  Vol.~43, American Astronomical Society Meeting Abstracts 217, 222.02--+

\bibitem[{{De~Rosa} {et~al.}(2011){De~Rosa}, {Walter}, W., \& {Fan}}]{DERO++11}
{De~Rosa}, G., {Walter}, F., W., R.~H., \& {Fan}, X. 2011, in preparation

\bibitem[{{Fan}(1999)}]{FAN99}
{Fan}, X. 1999, \aj, 117, 2528

\bibitem[{{Fan}(2006)}]{FAN06}
---. 2006, \nar, 50, 665

\bibitem[{{Fan} {et~al.}(2001){Fan}, {Narayanan}, {Lupton}, {Strauss}, {Knapp},
  {Becker}, {White}, {Pentericci}, {Leggett}, {Haiman}, {Gunn}, {Ivezi{\'c}},
  {Schneider}, {Anderson}, {Brinkmann}, {Bahcall}, {Connolly}, {Csabai}, {Doi},
  {Fukugita}, {Geballe}, {Grebel}, {Harbeck}, {Hennessy}, {Lamb}, {Miknaitis},
  {Munn}, {Nichol}, {Okamura}, {Pier}, {Prada}, {Richards}, {Szalay}, \&
  {York}}]{FAN++01}
{Fan}, X., {Narayanan}, V.~K., {Lupton}, R.~H., {Strauss}, M.~A., {Knapp},
  G.~R., {Becker}, R.~H., {White}, R.~L., {Pentericci}, L., {Leggett}, S.~K.,
  {Haiman}, Z., {Gunn}, J.~E., {Ivezi{\'c}}, {\v Z}., {Schneider}, D.~P.,
  {Anderson}, S.~F., {Brinkmann}, J., {Bahcall}, N.~A., {Connolly}, A.~J.,
  {Csabai}, I., {Doi}, M., {Fukugita}, M., {Geballe}, T., {Grebel}, E.~K.,
  {Harbeck}, D., {Hennessy}, G., {Lamb}, D.~Q., {Miknaitis}, G., {Munn}, J.~A.,
  {Nichol}, R., {Okamura}, S., {Pier}, J.~R., {Prada}, F., {Richards}, G.~T.,
  {Szalay}, A., \& {York}, D.~G. 2001, \aj, 122, 2833

\bibitem[{{Fan} {et~al.}(2002){Fan}, {Narayanan}, {Strauss}, {White}, {Becker},
  {Pentericci}, \& {Rix}}]{FAN++02}
{Fan}, X., {Narayanan}, V.~K., {Strauss}, M.~A., {White}, R.~L., {Becker},
  R.~H., {Pentericci}, L., \& {Rix}, H. 2002, \aj, 123, 1247

\bibitem[{{Fan} {et~al.}(2006){Fan}, {Strauss}, {Becker}, {White}, {Gunn},
  {Knapp}, {Richards}, {Schneider}, {Brinkmann}, \& {Fukugita}}]{FAN++06}
{Fan}, X., {Strauss}, M.~A., {Becker}, R.~H., {White}, R.~L., {Gunn}, J.~E.,
  {Knapp}, G.~R., {Richards}, G.~T., {Schneider}, D.~P., {Brinkmann}, J., \&
  {Fukugita}, M. 2006, \aj, 132, 117

\bibitem[{{Fan} {et~al.}(2003){Fan}, {Strauss}, {Schneider}, {Becker}, {White},
  {Haiman}, {Gregg}, {Pentericci}, {Grebel}, {Narayanan}, {Loh}, {Richards},
  {Gunn}, {Lupton}, {Knapp}, {Ivezi{\'c}}, {Brandt}, {Collinge}, {Hao},
  {Harbeck}, {Prada}, {Schaye}, {Strateva}, {Zakamska}, {Anderson},
  {Brinkmann}, {Bahcall}, {Lamb}, {Okamura}, {Szalay}, \& {York}}]{FAN++03}
{Fan}, X., {Strauss}, M.~A., {Schneider}, D.~P., {Becker}, R.~H., {White},
  R.~L., {Haiman}, Z., {Gregg}, M., {Pentericci}, L., {Grebel}, E.~K.,
  {Narayanan}, V.~K., {Loh}, Y.-S., {Richards}, G.~T., {Gunn}, J.~E., {Lupton},
  R.~H., {Knapp}, G.~R., {Ivezi{\'c}}, {\v Z}., {Brandt}, W.~N., {Collinge},
  M., {Hao}, L., {Harbeck}, D., {Prada}, F., {Schaye}, J., {Strateva}, I.,
  {Zakamska}, N., {Anderson}, S., {Brinkmann}, J., {Bahcall}, N.~A., {Lamb},
  D.~Q., {Okamura}, S., {Szalay}, A., \& {York}, D.~G. 2003, \aj, 125, 1649

\bibitem[{{Freudling} {et~al.}(2003){Freudling}, {Corbin}, \&
  {Korista}}]{FREU++03}
{Freudling}, W., {Corbin}, M.~R., \& {Korista}, K.~T. 2003, \apjl, 587, L67

\bibitem[{{Fukugita} {et~al.}(1996){Fukugita}, {Ichikawa}, {Gunn}, {Doi},
  {Shimasaku}, \& {Schneider}}]{FUJI++96}
{Fukugita}, M., {Ichikawa}, T., {Gunn}, J.~E., {Doi}, M., {Shimasaku}, K., \&
  {Schneider}, D.~P. 1996, \aj, 111, 1748

\bibitem[{{Greiner} {et~al.}(2008){Greiner}, {Bornemann}, {Clemens}, {Deuter},
  {Hasinger}, {Honsberg}, {Huber}, {Huber}, {Krauss}, {Kr{\"u}hler},
  {K{\"u}pc{\"u} Yolda{\c s}}, {Mayer-Hasselwander}, {Mican}, {Primak},
  {Schrey}, {Steiner}, {Szokoly}, {Th{\"o}ne}, {Yolda{\c s}}, {Klose}, {Laux},
  \& {Winkler}}]{GREI++08}
{Greiner}, J., {Bornemann}, W., {Clemens}, C., {Deuter}, M., {Hasinger}, G.,
  {Honsberg}, M., {Huber}, H., {Huber}, S., {Krauss}, M., {Kr{\"u}hler}, T.,
  {K{\"u}pc{\"u} Yolda{\c s}}, A., {Mayer-Hasselwander}, H., {Mican}, B.,
  {Primak}, N., {Schrey}, F., {Steiner}, I., {Szokoly}, G., {Th{\"o}ne}, C.~C.,
  {Yolda{\c s}}, A., {Klose}, S., {Laux}, U., \& {Winkler}, J. 2008, \pasp,
  120, 405

\bibitem[{{Gunn} \& {Peterson}(1965)}]{G&P65}
{Gunn}, J.~E. \& {Peterson}, B.~A. 1965, \apj, 142, 1633

\bibitem[{{Hewett} {et~al.}(2006){Hewett}, {Warren}, {Leggett}, \&
  {Hodgkin}}]{HEWE++06}
{Hewett}, P.~C., {Warren}, S.~J., {Leggett}, S.~K., \& {Hodgkin}, S.~T. 2006,
  \mnras, 367, 454

\bibitem[{{Jiang} {et~al.}(2008){Jiang}, {Fan}, {Annis}, {Becker}, {White},
  {Chiu}, {Lin}, {Lupton}, {Richards}, {Strauss}, {Jester}, \&
  {Schneider}}]{JIAN++08}
{Jiang}, L., {Fan}, X., {Annis}, J., {Becker}, R.~H., {White}, R.~L., {Chiu},
  K., {Lin}, H., {Lupton}, R.~H., {Richards}, G.~T., {Strauss}, M.~A.,
  {Jester}, S., \& {Schneider}, D.~P. 2008, \aj, 135, 1057

\bibitem[{{Jiang} {et~al.}(2007){Jiang}, {Fan}, {Vestergaard}, {Kurk},
  {Walter}, {Kelly}, \& {Strauss}}]{JIAN++07}
{Jiang}, L., {Fan}, X., {Vestergaard}, M., {Kurk}, J.~D., {Walter}, F.,
  {Kelly}, B.~C., \& {Strauss}, M.~A. 2007, \aj, 134, 1150

\bibitem[{{Kaiser} {et~al.}(2002){Kaiser}, {Aussel}, {Burke}, {Boesgaard},
  {Chambers}, {Chun}, {Heasley}, {Hodapp}, {Hunt}, {Jedicke}, {Jewitt},
  {Kudritzki}, {Luppino}, {Maberry}, {Magnier}, {Monet}, {Onaka}, {Pickles},
  {Rhoads}, {Simon}, {Szalay}, {Szapudi}, {Tholen}, {Tonry}, {Waterson}, \&
  {Wick}}]{KAIS++02}
{Kaiser}, N., {Aussel}, H., {Burke}, B.~E., {Boesgaard}, H., {Chambers}, K.,
  {Chun}, M.~R., {Heasley}, J.~N., {Hodapp}, K., {Hunt}, B., {Jedicke}, R.,
  {Jewitt}, D., {Kudritzki}, R., {Luppino}, G.~A., {Maberry}, M., {Magnier},
  E., {Monet}, D.~G., {Onaka}, P.~M., {Pickles}, A.~J., {Rhoads}, P.~H.~H.,
  {Simon}, T., {Szalay}, A., {Szapudi}, I., {Tholen}, D.~J., {Tonry}, J.~L.,
  {Waterson}, M., \& {Wick}, J. 2002, in Society of Photo-Optical
  Instrumentation Engineers (SPIE) Conference Series, Vol. 4836, Society of
  Photo-Optical Instrumentation Engineers (SPIE) Conference Series, ed.
  {J.~A.~Tyson \& S.~Wolff}, 154--164

\bibitem[{{Kaiser} {et~al.}(2010){Kaiser}, {Burgett}, {Chambers}, {Denneau},
  {Heasley}, {Jedicke}, {Magnier}, {Morgan}, {Onaka}, \& {Tonry}}]{KAIS++10}
{Kaiser}, N., {Burgett}, W., {Chambers}, K., {Denneau}, L., {Heasley}, J.,
  {Jedicke}, R., {Magnier}, E., {Morgan}, J., {Onaka}, P., \& {Tonry}, J. 2010,
  in Society of Photo-Optical Instrumentation Engineers (SPIE) Conference
  Series, Vol. 7733, Society of Photo-Optical Instrumentation Engineers (SPIE)
  Conference Series

\bibitem[{{Kembhavi} \& {Narlikar}(1999)}]{K&N99}
{Kembhavi}, A.~K. \& {Narlikar}, J.~V. 1999, {Quasars and active galactic
  nuclei : an introduction}, ed. {Kembhavi, A.~K.~\& Narlikar, J.~V.}
  ({Cambridge University Press})

\bibitem[{{Kurk} {et~al.}(2009){Kurk}, {Walter}, {Fan}, {Jiang}, {Jester},
  {Rix}, \& {Riechers}}]{KURK++09}
{Kurk}, J.~D., {Walter}, F., {Fan}, X., {Jiang}, L., {Jester}, S., {Rix}, H.,
  \& {Riechers}, D.~A. 2009, \apj, 702, 833

\bibitem[{{Kurk} {et~al.}(2007){Kurk}, {Walter}, {Riechers}, {Rix}, {Wagner},
  {Pentericci}, \& {Fan}}]{KURK++07}
{Kurk}, J.~D., {Walter}, F., {Riechers}, D., {Rix}, H.-W., {Wagner}, S.,
  {Pentericci}, L., \& {Fan}, X. 2007, Highlights of Astronomy, 14, 257

\bibitem[{{Lang} {et~al.}(2010){Lang}, {Hogg}, {Mierle}, {Blanton}, \&
  {Roweis}}]{LANG++10}
{Lang}, D., {Hogg}, D.~W., {Mierle}, K., {Blanton}, M., \& {Roweis}, S. 2010,
  \aj, 139, 1782

\bibitem[{{Lawrence} {et~al.}(2007){Lawrence}, {Warren}, {Almaini}, {Edge},
  {Hambly}, {Jameson}, {Lucas}, {Casali}, {Adamson}, {Dye}, {Emerson},
  {Foucaud}, {Hewett}, {Hirst}, {Hodgkin}, {Irwin}, {Lodieu}, {McMahon},
  {Simpson}, {Smail}, {Mortlock}, \& {Folger}}]{LAWR++07}
{Lawrence}, A., {Warren}, S.~J., {Almaini}, O., {Edge}, A.~C., {Hambly}, N.~C.,
  {Jameson}, R.~F., {Lucas}, P., {Casali}, M., {Adamson}, A., {Dye}, S.,
  {Emerson}, J.~P., {Foucaud}, S., {Hewett}, P., {Hirst}, P., {Hodgkin}, S.~T.,
  {Irwin}, M.~J., {Lodieu}, N., {McMahon}, R.~G., {Simpson}, C., {Smail}, I.,
  {Mortlock}, D., \& {Folger}, M. 2007, \mnras, 379, 1599

\bibitem[{{Mortlock} {et~al.}(2008){Mortlock}, {Patel}, {Warren}, {Venemans},
  {McMahon}, {Hewett}, {Simpson}, \& {Sharp}}]{MORT++08}
{Mortlock}, D.~J., {Patel}, M., {Warren}, S.~J., {Venemans}, B.~P., {McMahon},
  R.~G., {Hewett}, P., {Simpson}, C., \& {Sharp}, R.~G. 2008, ArXiv e-prints

\bibitem[{{Mortlock} {et~al.}(2011){Mortlock}, {Warren}, {Venemans}, {Patel},
  {Hewett}, {McMahon}, {Simpson}, {Theuns}, {Gonz{\'a}les-Solares}, {Adamson},
  {Dye}, {Hambly}, {Hirst}, {Irwin}, {Kuiper}, {Lawrence}, \&
  {R{\"o}ttgering}}]{MORT++11b}
{Mortlock}, D.~J., {Warren}, S.~J., {Venemans}, B.~P., {Patel}, M., {Hewett},
  P.~C., {McMahon}, R.~G., {Simpson}, C., {Theuns}, T., {Gonz{\'a}les-Solares},
  E.~A., {Adamson}, A., {Dye}, S., {Hambly}, N.~C., {Hirst}, P., {Irwin},
  M.~J., {Kuiper}, E., {Lawrence}, A., \& {R{\"o}ttgering}, H.~J.~A. 2011,
  \nat, 474, 616

\bibitem[{{Price} {et~al.}(2007){Price}, {Chambers}, {Jester}, \&
  {Walter}}]{PRIC++07}
{Price}, P.~A., {Chambers}, K.~C., {Jester}, S., \& {Walter}, F. 2007, in
  Bulletin of the American Astronomical Society, Vol.~38, American Astronomical
  Society Meeting Abstracts, 807--+

\bibitem[{{Rees}(1984)}]{REES84}
{Rees}, M.~J. 1984, \araa, 22, 471

\bibitem[{{Richards} {et~al.}(2006){Richards}, {Lacy}, {Storrie-Lombardi},
  {Hall}, {Gallagher}, {Hines}, {Fan}, {Papovich}, {Vanden Berk}, {Trammell},
  {Schneider}, {Vestergaard}, {York}, {Jester}, {Anderson}, {Budav{\'a}ri}, \&
  {Szalay}}]{RICH++06}
{Richards}, G.~T., {Lacy}, M., {Storrie-Lombardi}, L.~J., {Hall}, P.~B.,
  {Gallagher}, S.~C., {Hines}, D.~C., {Fan}, X., {Papovich}, C., {Vanden Berk},
  D.~E., {Trammell}, G.~B., {Schneider}, D.~P., {Vestergaard}, M., {York},
  D.~G., {Jester}, S., {Anderson}, S.~F., {Budav{\'a}ri}, T., \& {Szalay},
  A.~S. 2006, \apjs, 166, 470

\bibitem[{{Skrutskie} {et~al.}(2006){Skrutskie}, {Cutri}, {Stiening},
  {Weinberg}, {Schneider}, {Carpenter}, {Beichman}, {Capps}, {Chester},
  {Elias}, {Huchra}, {Liebert}, {Lonsdale}, {Monet}, {Price}, {Seitzer},
  {Jarrett}, {Kirkpatrick}, {Gizis}, {Howard}, {Evans}, {Fowler}, {Fullmer},
  {Hurt}, {Light}, {Kopan}, {Marsh}, {McCallon}, {Tam}, {Van Dyk}, \&
  {Wheelock}}]{SKRU++06}
{Skrutskie}, M.~F., {Cutri}, R.~M., {Stiening}, R., {Weinberg}, M.~D.,
  {Schneider}, S., {Carpenter}, J.~M., {Beichman}, C., {Capps}, R., {Chester},
  T., {Elias}, J., {Huchra}, J., {Liebert}, J., {Lonsdale}, C., {Monet}, D.~G.,
  {Price}, S., {Seitzer}, P., {Jarrett}, T., {Kirkpatrick}, J.~D., {Gizis},
  J.~E., {Howard}, E., {Evans}, T., {Fowler}, J., {Fullmer}, L., {Hurt}, R.,
  {Light}, R., {Kopan}, E.~L., {Marsh}, K.~A., {McCallon}, H.~L., {Tam}, R.,
  {Van Dyk}, S., \& {Wheelock}, S. 2006, \aj, 131, 1163

\bibitem[{{Trump} {et~al.}(2006){Trump}, {Hall}, {Reichard}, {Richards},
  {Schneider}, {Vanden Berk}, {Knapp}, {Anderson}, {Fan}, {Brinkman},
  {Kleinman}, \& {Nitta}}]{TRUM++06}
{Trump}, J.~R., {Hall}, P.~B., {Reichard}, T.~A., {Richards}, G.~T.,
  {Schneider}, D.~P., {Vanden Berk}, D.~E., {Knapp}, G.~R., {Anderson}, S.~F.,
  {Fan}, X., {Brinkman}, J., {Kleinman}, S.~J., \& {Nitta}, A. 2006, \apjs,
  165, 1

\bibitem[{{Vanden Berk} {et~al.}(2001){Vanden Berk}, {Richards}, {Bauer},
  {Strauss}, {Schneider}, {Heckman}, {York}, {Hall}, {Fan}, {Knapp},
  {Anderson}, {Annis}, {Bahcall}, {Bernardi}, {Briggs}, {Brinkmann}, {Brunner},
  {Burles}, {Carey}, {Castander}, {Connolly}, {Crocker}, {Csabai}, {Doi},
  {Finkbeiner}, {Friedman}, {Frieman}, {Fukugita}, {Gunn}, {Hennessy},
  {Ivezi{\'c}}, {Kent}, {Kunszt}, {Lamb}, {Leger}, {Long}, {Loveday}, {Lupton},
  {Meiksin}, {Merelli}, {Munn}, {Newberg}, {Newcomb}, {Nichol}, {Owen}, {Pier},
  {Pope}, {Rockosi}, {Schlegel}, {Siegmund}, {Smee}, {Snir}, {Stoughton},
  {Stubbs}, {SubbaRao}, {Szalay}, {Szokoly}, {Tremonti}, {Uomoto}, {Waddell},
  {Yanny}, \& {Zheng}}]{VAND++01}
{Vanden Berk}, D.~E., {Richards}, G.~T., {Bauer}, A., {Strauss}, M.~A.,
  {Schneider}, D.~P., {Heckman}, T.~M., {York}, D.~G., {Hall}, P.~B., {Fan},
  X., {Knapp}, G.~R., {Anderson}, S.~F., {Annis}, J., {Bahcall}, N.~A.,
  {Bernardi}, M., {Briggs}, J.~W., {Brinkmann}, J., {Brunner}, R., {Burles},
  S., {Carey}, L., {Castander}, F.~J., {Connolly}, A.~J., {Crocker}, J.~H.,
  {Csabai}, I., {Doi}, M., {Finkbeiner}, D., {Friedman}, S., {Frieman}, J.~A.,
  {Fukugita}, M., {Gunn}, J.~E., {Hennessy}, G.~S., {Ivezi{\'c}}, {\v Z}.,
  {Kent}, S., {Kunszt}, P.~Z., {Lamb}, D.~Q., {Leger}, R.~F., {Long}, D.~C.,
  {Loveday}, J., {Lupton}, R.~H., {Meiksin}, A., {Merelli}, A., {Munn}, J.~A.,
  {Newberg}, H.~J., {Newcomb}, M., {Nichol}, R.~C., {Owen}, R., {Pier}, J.~R.,
  {Pope}, A., {Rockosi}, C.~M., {Schlegel}, D.~J., {Siegmund}, W.~A., {Smee},
  S., {Snir}, Y., {Stoughton}, C., {Stubbs}, C., {SubbaRao}, M., {Szalay},
  A.~S., {Szokoly}, G.~P., {Tremonti}, C., {Uomoto}, A., {Waddell}, P.,
  {Yanny}, B., \& {Zheng}, W. 2001, \aj, 122, 549

\bibitem[{{Venemans}(2007)}]{VENE07}
{Venemans}, B.~P. 2007, in Astronomical Society of the Pacific Conference
  Series, Vol. 379, Cosmic Frontiers, ed. {N.~Metcalfe \& T.~Shanks}, 43--+

\bibitem[{{Villata} {et~al.}(2006){Villata}, {Raiteri}, {Balonek}, {Aller},
  {Jorstad}, {Kurtanidze}, {Nicastro}, {Nilsson}, {Aller}, {Arai}, {Arkharov},
  {Bach}, {Ben{\'{\i}}tez}, {Berdyugin}, {Buemi}, {B{\"o}ttcher}, {Carosati},
  {Casas}, {Caulet}, {Chen}, {Chiang}, {Chou}, {Ciprini}, {Coloma}, {di Rico},
  {D{\'{\i}}az}, {Efimova}, {Forsyth}, {Frasca}, {Fuhrmann}, {Gadway}, {Gupta},
  {Hagen-Thorn}, {Harvey}, {Heidt}, {Hernandez-Toledo}, {Hroch}, {Hu}, {Hudec},
  {Ibrahimov}, {Imada}, {Kamata}, {Kato}, {Katsuura}, {Konstantinova},
  {Kopatskaya}, {Kotaka}, {Kovalev}, {Kovalev}, {Krichbaum}, {Kubota},
  {Kurosaki}, {Lanteri}, {Larionov}, {Larionova}, {Laurikainen}, {Lee}, {Leto},
  {L{\"a}hteenm{\"a}ki}, {L{\'o}pez-Cruz}, {Marilli}, {Marscher}, {McHardy},
  {Mondal}, {Mullan}, {Napoleone}, {Nikolashvili}, {Ohlert}, {Postnikov},
  {Pursimo}, {Ragni}, {Ros}, {Sadakane}, {Sadun}, {Savolainen}, {Sergeeva},
  {Sigua}, {Sillanp{\"a}{\"a}}, {Sixtova}, {Sumitomo}, {Takalo},
  {Ter{\"a}sranta}, {Tornikoski}, {Trigilio}, {Umana}, {Volvach}, {Voss}, \&
  {Wortel}}]{VILL++06}
{Villata}, M., {Raiteri}, C.~M., {Balonek}, T.~J., {Aller}, M.~F., {Jorstad},
  S.~G., {Kurtanidze}, O.~M., {Nicastro}, F., {Nilsson}, K., {Aller}, H.~D.,
  {Arai}, A., {Arkharov}, A., {Bach}, U., {Ben{\'{\i}}tez}, E., {Berdyugin},
  A., {Buemi}, C.~S., {B{\"o}ttcher}, M., {Carosati}, D., {Casas}, R.,
  {Caulet}, A., {Chen}, W.~P., {Chiang}, P., {Chou}, Y., {Ciprini}, S.,
  {Coloma}, J.~M., {di Rico}, G., {D{\'{\i}}az}, C., {Efimova}, N.~V.,
  {Forsyth}, C., {Frasca}, A., {Fuhrmann}, L., {Gadway}, B., {Gupta}, S.,
  {Hagen-Thorn}, V.~A., {Harvey}, J., {Heidt}, J., {Hernandez-Toledo}, H.,
  {Hroch}, F., {Hu}, C., {Hudec}, R., {Ibrahimov}, M.~A., {Imada}, A.,
  {Kamata}, M., {Kato}, T., {Katsuura}, M., {Konstantinova}, T., {Kopatskaya},
  E., {Kotaka}, D., {Kovalev}, Y.~Y., {Kovalev}, Y.~A., {Krichbaum}, T.~P.,
  {Kubota}, K., {Kurosaki}, M., {Lanteri}, L., {Larionov}, V.~M., {Larionova},
  L., {Laurikainen}, E., {Lee}, C., {Leto}, P., {L{\"a}hteenm{\"a}ki}, A.,
  {L{\'o}pez-Cruz}, O., {Marilli}, E., {Marscher}, A.~P., {McHardy}, I.~M.,
  {Mondal}, S., {Mullan}, B., {Napoleone}, N., {Nikolashvili}, M.~G., {Ohlert},
  J.~M., {Postnikov}, S., {Pursimo}, T., {Ragni}, M., {Ros}, J.~A., {Sadakane},
  K., {Sadun}, A.~C., {Savolainen}, T., {Sergeeva}, E.~A., {Sigua}, L.~A.,
  {Sillanp{\"a}{\"a}}, A., {Sixtova}, L., {Sumitomo}, N., {Takalo}, L.~O.,
  {Ter{\"a}sranta}, H., {Tornikoski}, M., {Trigilio}, C., {Umana}, G.,
  {Volvach}, A., {Voss}, B., \& {Wortel}, S. 2006, \aap, 453, 817

\bibitem[{{Willott} {et~al.}(2007){Willott}, {Delorme}, {Omont}, {Bergeron},
  {Delfosse}, {Forveille}, {Albert}, {Reyl{\'e}}, {Hill}, {Gully-Santiago},
  {Vinten}, {Crampton}, {Hutchings}, {Schade}, {Simard}, {Sawicki}, {Beelen},
  \& {Cox}}]{WILL++07}
{Willott}, C.~J., {Delorme}, P., {Omont}, A., {Bergeron}, J., {Delfosse}, X.,
  {Forveille}, T., {Albert}, L., {Reyl{\'e}}, C., {Hill}, G.~J.,
  {Gully-Santiago}, M., {Vinten}, P., {Crampton}, D., {Hutchings}, J.~B.,
  {Schade}, D., {Simard}, L., {Sawicki}, M., {Beelen}, A., \& {Cox}, P. 2007,
  \aj, 134, 2435

\bibitem[{{Willott} {et~al.}(2010){Willott}, {Delorme}, {Reyl{\'e}}, {Albert},
  {Bergeron}, {Crampton}, {Delfosse}, {Forveille}, {Hutchings}, {McLure},
  {Omont}, \& {Schade}}]{WILL++10}
{Willott}, C.~J., {Delorme}, P., {Reyl{\'e}}, C., {Albert}, L., {Bergeron}, J.,
  {Crampton}, D., {Delfosse}, X., {Forveille}, T., {Hutchings}, J.~B.,
  {McLure}, R.~J., {Omont}, A., \& {Schade}, D. 2010, \aj, 139, 906

\bibitem[{{York} {et~al.}(2000){York}, {Adelman}, {Anderson}, {Anderson},
  {Annis}, {Bahcall}, {Bakken}, {Barkhouser}, {Bastian}, {Berman}, {Boroski},
  {Bracker}, {Briegel}, {Briggs}, {Brinkmann}, {Brunner}, {Burles}, {Carey},
  {Carr}, {Castander}, {Chen}, {Colestock}, {Connolly}, {Crocker}, {Csabai},
  {Czarapata}, {Davis}, {Doi}, {Dombeck}, {Eisenstein}, {Ellman}, {Elms},
  {Evans}, {Fan}, {Federwitz}, {Fiscelli}, {Friedman}, {Frieman}, {Fukugita},
  {Gillespie}, {Gunn}, {Gurbani}, {de Haas}, {Haldeman}, {Harris}, {Hayes},
  {Heckman}, {Hennessy}, {Hindsley}, {Holm}, {Holmgren}, {Huang}, {Hull},
  {Husby}, {Ichikawa}, {Ichikawa}, {Ivezi{\'c}}, {Kent}, {Kim}, {Kinney},
  {Klaene}, {Kleinman}, {Kleinman}, {Knapp}, {Korienek}, {Kron}, {Kunszt},
  {Lamb}, {Lee}, {Leger}, {Limmongkol}, {Lindenmeyer}, {Long}, {Loomis},
  {Loveday}, {Lucinio}, {Lupton}, {MacKinnon}, {Mannery}, {Mantsch}, {Margon},
  {McGehee}, {McKay}, {Meiksin}, {Merelli}, {Monet}, {Munn}, {Narayanan},
  {Nash}, {Neilsen}, {Neswold}, {Newberg}, {Nichol}, {Nicinski}, {Nonino},
  {Okada}, {Okamura}, {Ostriker}, {Owen}, {Pauls}, {Peoples}, {Peterson},
  {Petravick}, {Pier}, {Pope}, {Pordes}, {Prosapio}, {Rechenmacher}, {Quinn},
  {Richards}, {Richmond}, {Rivetta}, {Rockosi}, {Ruthmansdorfer}, {Sandford},
  {Schlegel}, {Schneider}, {Sekiguchi}, {Sergey}, {Shimasaku}, {Siegmund},
  {Smee}, {Smith}, {Snedden}, {Stone}, {Stoughton}, {Strauss}, {Stubbs},
  {SubbaRao}, {Szalay}, {Szapudi}, {Szokoly}, {Thakar}, {Tremonti}, {Tucker},
  {Uomoto}, {Vanden Berk}, {Vogeley}, {Waddell}, {Wang}, {Watanabe},
  {Weinberg}, {Yanny}, \& {Yasuda}}]{YORK++00}
{York}, D.~G., {Adelman}, J., {Anderson}, Jr., J.~E., {Anderson}, S.~F.,
  {Annis}, J., {Bahcall}, N.~A., {Bakken}, J.~A., {Barkhouser}, R., {Bastian},
  S., {Berman}, E., {Boroski}, W.~N., {Bracker}, S., {Briegel}, C., {Briggs},
  J.~W., {Brinkmann}, J., {Brunner}, R., {Burles}, S., {Carey}, L., {Carr},
  M.~A., {Castander}, F.~J., {Chen}, B., {Colestock}, P.~L., {Connolly}, A.~J.,
  {Crocker}, J.~H., {Csabai}, I., {Czarapata}, P.~C., {Davis}, J.~E., {Doi},
  M., {Dombeck}, T., {Eisenstein}, D., {Ellman}, N., {Elms}, B.~R., {Evans},
  M.~L., {Fan}, X., {Federwitz}, G.~R., {Fiscelli}, L., {Friedman}, S.,
  {Frieman}, J.~A., {Fukugita}, M., {Gillespie}, B., {Gunn}, J.~E., {Gurbani},
  V.~K., {de Haas}, E., {Haldeman}, M., {Harris}, F.~H., {Hayes}, J.,
  {Heckman}, T.~M., {Hennessy}, G.~S., {Hindsley}, R.~B., {Holm}, S.,
  {Holmgren}, D.~J., {Huang}, C., {Hull}, C., {Husby}, D., {Ichikawa}, S.,
  {Ichikawa}, T., {Ivezi{\'c}}, {\v Z}., {Kent}, S., {Kim}, R.~S.~J., {Kinney},
  E., {Klaene}, M., {Kleinman}, A.~N., {Kleinman}, S., {Knapp}, G.~R.,
  {Korienek}, J., {Kron}, R.~G., {Kunszt}, P.~Z., {Lamb}, D.~Q., {Lee}, B.,
  {Leger}, R.~F., {Limmongkol}, S., {Lindenmeyer}, C., {Long}, D.~C., {Loomis},
  C., {Loveday}, J., {Lucinio}, R., {Lupton}, R.~H., {MacKinnon}, B.,
  {Mannery}, E.~J., {Mantsch}, P.~M., {Margon}, B., {McGehee}, P., {McKay},
  T.~A., {Meiksin}, A., {Merelli}, A., {Monet}, D.~G., {Munn}, J.~A.,
  {Narayanan}, V.~K., {Nash}, T., {Neilsen}, E., {Neswold}, R., {Newberg},
  H.~J., {Nichol}, R.~C., {Nicinski}, T., {Nonino}, M., {Okada}, N., {Okamura},
  S., {Ostriker}, J.~P., {Owen}, R., {Pauls}, A.~G., {Peoples}, J., {Peterson},
  R.~L., {Petravick}, D., {Pier}, J.~R., {Pope}, A., {Pordes}, R., {Prosapio},
  A., {Rechenmacher}, R., {Quinn}, T.~R., {Richards}, G.~T., {Richmond}, M.~W.,
  {Rivetta}, C.~H., {Rockosi}, C.~M., {Ruthmansdorfer}, K., {Sandford}, D.,
  {Schlegel}, D.~J., {Schneider}, D.~P., {Sekiguchi}, M., {Sergey}, G.,
  {Shimasaku}, K., {Siegmund}, W.~A., {Smee}, S., {Smith}, J.~A., {Snedden},
  S., {Stone}, R., {Stoughton}, C., {Strauss}, M.~A., {Stubbs}, C., {SubbaRao},
  M., {Szalay}, A.~S., {Szapudi}, I., {Szokoly}, G.~P., {Thakar}, A.~R.,
  {Tremonti}, C., {Tucker}, D.~L., {Uomoto}, A., {Vanden Berk}, D., {Vogeley},
  M.~S., {Waddell}, P., {Wang}, S., {Watanabe}, M., {Weinberg}, D.~H., {Yanny},
  B., \& {Yasuda}, N. 2000, \aj, 120, 1579

\end{thebibliography}

\end{document}